\documentclass[12pt,epsfig]{article}
\def\baselinestretch{1.5}
\begin{document}
%\input{psfig.tex}
%\magnification=1200
\baselineskip=20 pt
\def\l{\lambda}
\def\L{\Lambda}
\def\b{\beta}
\def\mphi{m_{\phi}}
\def\dnul{\partial_{\nu}}
\def\dnuu{\partial^{\nu}}
\def\dmul{\partial_{\mu}}
\def\dmuu{\partial^{\mu}}
\def\eps{\epsilon}
\def\hphi{\hat{\phi}}
\def\vphi{\langle \phi \rangle}
\def\etamunu{\eta^{\mu\nu}}
\def\bfl{\begin{flushleft}}
\def\efl{\end{flushleft}}
\def\bea{\begin{eqnarray}}
\def\eea{\end{eqnarray}}
\begin{center}
{\huge\bf
   {$\rho$} parameter constraints on radion phenomenology and a
 lower bound on Higgs mass\\}  
 
\end{center}

\vskip 10pT
\begin{center}
{\large\sl \bf{Prasanta Das}~\footnote{E-mail: pdas@iitk.ac.in} 
}
\vskip  5pT
{\rm
Department of Physics, Indian Institute of Technology, \\
Kanpur 208 016, India.} \\
\end{center}

%\vskip 0.02in

\begin{center}
{\large\sl  \bf{Uma Mahanta}~\footnote{E-mail:mahanta@mri.ernet.in}
}
\vskip 5pT
{\rm
Mehta Research Institute, \\
Chhatnag Road, Jhusi
Allahabad-211019, India .}\\
\end{center}

\centerline{\bf Abstract}

In this paper we determine the contribution of a light stabilized
  radion to the
weak isospin violating $\rho$ parameter by using an ultraviolet
momentum cut off as the regulator. The LEP1 bound on $\rho^{new}$
is then used to derive constraints on the radion mass $\mphi$ and its vev
$\vphi$. Finally by using the beta function of the higgs self coupling
we have determined a lower bound on the higgs mass from the rho parameter
constraints on $\mphi$ and $\vphi$. Our results show that for 
$\mphi <$ 600 GeV the rho parameter bound on $m_h$ is stronger than the 
present direct bound from LEPII.

\vfill\eject

\centerline{\Large \bf Introduction}

Recently several attractive
proposals (\cite{ADD} \cite{RS}) based on theories in 
extra dimensions have been
put forward to explain the hierarchy problem. 
Among them the Randall-Sundrum (RS) model is particularly interesting
because it considers a five dimensional world based on the
following non-factorizable metric

\bea
ds^2= e^{-2k r_{c}|\theta|}{\eta_{\mu\nu}} dx^{\mu} dx^{\nu}- {r_c}^2 
d{\theta}^2
\eea

Here $r_c$ measures the size of the extra dimensions which is an ${S^1/Z_2}$
orbifold. $x^{\mu }$ are the coordinates of the four
dimensional space-time. $-\pi\le \theta \le \pi$ is the coordinate
 of the extra dimension
with $\theta$ and $-\theta$ identified. k is a mass parameter of the order
of the fundamental five dimensional Planck mass M. 
Two 3 branes are placed at the orbifold
fixed points $\theta =0$ (hidden brane) and $\theta =\pi$ (visible
brane). Randall and Sundrum showed that any field on the visible brane
with a fundamental 
 mass parameter $m_0$ gets an effective mass 

\bea
m = m_0 e^{-k r_c \pi}
\eea

due to the exponential warp factor. Therefore
 for $k r_c \approx 14$ the electroweak (EW) scale
is generated from the Planck scale by the warp factor.

In the Randall-Sundrum model $r_{c}$ is the vacuum expectation 
value (vev) of a 
massless scalar field T(x). The modulus was therefore not stabilized
by some dynamics. Goldberger and Wise {\cite{GW}} 
later showed how to generate a
potential for the modulus and stabilize it at the right value 
($k r_c$) that is
needed for solving the hierarchy problem without any excessive
fine tuning of the parameters.

In their original model  Randall and Sundrum  assumed  that SM fileds are 
localized on the visible brane located at $\theta =\pi$. 
However, small fluctuations of the modulus
field from its vev gives rise to non-trivial couplings  of 
the modulus field with the SM fields. Using such couplings the effect
of a light radion on low energy phenomenology has been studied
{\cite{CGRT}}.

In this report we shall 
determine the radion contribution to the weak isospin breaking
$\rho$ parameter.
The LEP1 data
imposes stringent constraints on new physics contribution to the $\rho$
parameter. We have therefore used the LEP1 bound on $\rho^{new}$  to put 
bounds on the two unknown parameters $m_{\phi}$ and $\vphi$.
Throughout our analysis the cut off $\Lambda$ will be assumed to be related
to the expansion parameter ${1\over \vphi }$ 
 of the non-renormalizable radion interaction to SM particles
by the naive dimensional analysis (NDA) estimate $\L =4\pi \vphi$
 {\cite{GM}}. In the RS model the cut-off $\L$ corresponds to the mass of
the lightest KK graviton mode.
The beta
function of the Higgs self-coupling is also modified in the presence of a
light stablized radion.  In this paper, we have used this beta function to
derive a lower bound on $m_h$ from the LEP1 bounds on $m_{\phi}$ 
and $\vphi$ .\\

\centerline{\bf Radion couplings to electroweak gauge bosons in unitary gauge}

In order to determine the radion contribution to the T parameter
 we need to determine the radion couplings
to the EW gauge bosons localized on the  visible brane. The 
relevant  radion couplings to the EW gauge bosons can be determined from the 
following action 

\vspace*{-0.25in}

\bfl
\bea
S = \int d^4 x {\sqrt{- g_{v}}}\left [ (D_{\mu}H )^{\dagger}(D_{\nu}H )
g^{\mu\nu}_{v} - \frac{1}{4} W^a_{\mu\nu} W^a_{\rho\sigma} g^{\mu\rho}_v
g^{\nu\sigma}_v - \frac{1}{4} B_{\mu\nu} B_{\rho\sigma} g^{\mu\rho}_v
g^{\nu\sigma}_v + {\mathcal{L}}_{gf} \right ] 
\eea
\efl

where ${\sqrt{- g_v}} = (\frac{\phi}{f})^4$ ,~ $g^{\mu\nu}_v =
(\frac{\phi (x)}{f})^{- 2} \eta^{\mu\nu}$

and

\bea
{\mathcal{L}}_{gf} = - \frac{1}{2 \xi} \left [ \partial_{\mu} W^{a}_{\nu}
~g^{\mu\nu}_v + i g_2~ \xi \left ( {H^{\prime}}^{\dagger}
{\frac{\tau_a}{2}}<H> -
<H^{\dagger}>{\frac{\tau_a}{2}} H^{\prime} \right ) \right ]^2 \noindent
\nonumber \\
- \frac{1}{2 \xi} \left [ \partial_{\mu} B_{\nu}~g^{\mu\nu}_v + i
g_1
~{\frac{\xi}{2}} \left ( {H^{\prime}}^{\dagger} <H> - <H^{\dagger}> H^{\prime}
\right ) \right ]^2 
\eea

The SM higgs field in unitary gauge is given by
$$
H = H^{\prime} + <H> =  
 \pmatrix{0\cr {\frac{v+h(x)}{\sqrt{2}}}\cr}
$$

Using the above expression of the higgs field it can be shown that
the gauge fixing Lagrangian $L_{gf}$
 vanishes in the unitary gauge ($\xi\rightarrow
\infty$).
Consider first the radion coupling to the K.E. of the gauge bosons. 
We have

\bea
\sqrt{- g_v}~ V_{\mu\nu}V_{\rho\sigma} g_v^{\mu\rho} g_v^{\nu\sigma} =
V_{\mu\nu} V_{\rho\sigma} \eta^{\mu\rho} \eta^{\nu\sigma}   
\eea

where $V_{\mu}=( W^a_{\mu}, B_{\mu})$. At the classical level the 
radion therefore does not couple to the gauge boson KE in four dimensions.
 Note that the Christoffel symbol
$\Gamma^{\lambda}_{\mu\nu}$ in the expression for the general covariant
derivative $D_{\mu}V_{\nu}$ does not contribute to the field strength
tensor of $W^{a}_{\nu}$ or $B_{\nu}$ because $\Gamma^{\lambda}_{\mu\nu}$
is symmetric in ($\mu,\nu$).
 
Consider next the radion coupling to the K.E. of the higgs boson
We have

$$\sqrt{-g_v}(D_{\mu}H)^+(D_{\nu}H)g_v^{\mu\nu} =
\left({\phi\over \vphi}\right)^2(D_{\mu}{\tilde
{H}})^+(D^{\mu}{\tilde{H}})$$

where  ${\tilde{H}} = H~ \left(\frac{\vphi}{f}\right)$ and
 ${\tilde{<H>}} =~ <H>~ \left(\frac{\vphi}{f}\right)$. 
In the following we shall assume 
that the higgs field and its vev
 has been properly rescaled as above and drop the
tilde sign. We then get

\bea
\sqrt{-g_v}(D_{\mu}H)^+(D_{\nu}H)g_v^{\mu\nu} =
i g_2 W^{a}_{\mu}\left[<H^{+}> \frac{\tau_a}{2} \partial^{\mu} H^{\prime}~
-~ \partial^{\mu} H^{\prime +} \frac{\tau_a}{2}<H>\right]
\left( \frac{\phi}{\vphi}\right)^2 \noindent
\nonumber \\
+ i {g_1\over 2}B_{\mu}\left[<H^{+}> \dmuu H^{\prime}- \dmuu H^{\prime+}
<H>\right] \left(\phi \over {\vphi}\right)^2 \noindent
\nonumber \\
+\left[ m^2_w W^{+}_{\mu}W^{-\mu} + {1\over 2}m^2_z Z_{\mu}Z^{\mu} \right]
\left[1+2{\hat{\phi}\over \vphi}+ {{\hat
{\phi}}^2
\over \vphi^2}~ +..\right]
\eea

In the unitary gauge the first two terms on the rhs of the above
expression vanishes, leaving only the gauge boson mass terms to couple
to radion fluctuations.
 The couplings of one and two radions to the EW
gauge bosons that are relevant for computing the radion contribution
to $\Pi^{\mu\nu}_{vv}(q)$ in unitary gauge can therefore be expressed
by the Feynman rules shown in Fig1(a) and Fig1(b).

The contribution of new physics to the vectorial isospin violating parameter
 $\rho$  {\cite{PT}} is given by

\bea
\rho^{new}=\alpha T^{new}={\Pi_{ww}(0)\over m^2_w}-{\Pi_{zz}(0)\over
m^2_z}
\eea

The function $\Pi_{vv}(q)$ is defined through the gauge boson self energy
tensor

\bea
i\Pi^{\mu\nu}_{vv}(q)=i\eta^{\mu\nu}\Pi_{vv}(q)-iq^{\mu}q^{\nu}
\tilde {\Pi}_{vv}(q)
\eea

The Feynman diagrams that give rise to radion contribution to
$\rho$ parameter in unitary gauge are shown in Fig 2.

Let $\Pi^{(1)}_{vv}(q)$
and $\Pi^{(2)}_{vv}(q)$ denote the  contributions arising
 from single and two radion vertices to $\Pi_{vv}(q)$. We then have

\bea
\Pi^{(1)}_{vv}(0) = -{m_v^2\over 16\pi^2\vphi^2}\left(\L^2-\mphi^2 \ln {\L^2
\over \mphi^2})\right)\noindent
\nonumber \\
-{m_v^4\over 16\pi^2}[3\ln {\L^2\over\mphi^2}-3{m^2_v\over \mphi^2 -m^2_v}
\ln {\mphi^2\over m_v^2}
\eea

and 

\bea
\Pi^{(2)}_{vv}(0)={m_v^2\over 16\pi^2\vphi^2}(\L^2 -\mphi^2\ln{\L^2\over
\mphi^2})
\eea

 It is clear from the above that 
$\Pi^{(2)}_{vv}(q)$ will not contribute to the $\rho$ parameter
since ${\Pi^{(2)}_{vv}(0)\over m_v^2}$ is independent of $m_v$.
 Radion contribution to $\rho^{new}$
therefore arises only from $\Pi^{(1)}_{vv}(q)$.
 We would like to note at this
point that since the $\hat {\phi}VV$ 
 coupling is proportional to $m^2_v$, the radion tadpole diagrams 
do not contribute to the $\rho$ parameter.
We also find that although $\Pi^{(1)}_{vv}(0)$ and $\Pi^{(2)}_{vv}(0)$
are individually quadratically divergent the sum $\Pi_{vv}(0)$
is only log divergent. This cancellation of quadratic divergence
is a consequence of gauge symmetry which protects gauge boson masses
from receiving large power corrections.

\centerline{\bf Radion contribution to the $\rho$ parameter}

The radion contribution to the $\rho$ parameter is therefore given by

\bea
\rho^{new} = {m_w^2\over 16\pi^2\vphi^2}\left[
- 3 ln{\L^2\over \mphi^2}+{3m_w^2\over {\mphi^2-m_w^2}}~ln{\mphi^2\over
m_w^2}\right] \noindent 
\nonumber \\
- {m_z^2\over 16\pi^2\vphi^2}\left[
- 3 ln{\L^2\over \mphi^2}+{3m_z^2\over {\mphi^2-m_z^2}}~ln{\mphi^2\over
m_z^2}\right]
\eea

Note that the above expression for  $\rho^{new}$ diverges
logarithmically with the cut off.
The cut off dependence of the radion contribution to the rho parameter
is easily understood. It arises from using the non-renormalizable
dimension five operator $(D_{\mu}H)^+(D^{\mu}H){\hat{\phi}\over \vphi}$
in the calculation of $\Pi^1_{vv}(0)$.
Secondly the radion contribution to $\rho^{new}$
depends on three unknown
parameters: the cut off $\L$, $\mphi$ and $\vphi$. 
Naive dimensional analysis can however be used to relate $\L$ to the
expansion parameter ${1\over \vphi}$ through $\L = 4\pi \vphi$.
 Physically it means that the radion effective field
theory would become non-perturbative and radion induced radiative
corrections would become very large above the ultraviolet
 scale $4\pi \vphi$ thereby implying a breakdown of the low energy
effective theory. 
 The NDA estimate of the cut off is known to work
quite well for
estimating chiral loops that arises in dealing with Chiral Lagrangian.
  Further Luty et al has shown
that it gives reliable estimates for extra dimensional gravity also.
We shall assume it to hold good for the radion effective field theory
also. This reduces the
dependence of $\rho^{new}$ to two unknowns only: $\mphi$ and $\vphi$.
The LEPI bounds on $\rho^{new}$ can therefore be used to impose stringent 
constraints on $\mphi$ and $\vphi$ {\cite{HMZ}}.
Finally
the radion contribution to the T parameter 
must be a gauge invariant quantity, since the radion is a gauge singlet.
 Therefore although the 
calculations presented in this paper were done in unitary gauge, the final
answer given by eqn (11) must be independent of this gauge choice. 

\bfl
{\Large \bf { $\rho$ parameter  bounds on $\mphi$ and $\vphi$}}
\efl

The present value of the T parameter is given  by {\cite{EPJ}}
$T=-0.10 \pm 0.14~ (0.09)$. In fig 3 we have shown the contour plots in
$\mphi$
vs $\vphi$ plane for T=0.04 and T =0.18. The first value corresponds to 
+1$\sigma$ deviation and the second value to 2$\sigma $ deviation from 
the central value. We have chosen positive values of T only since the
 radion contribution to T is positive for $\L \gg \mphi$. The region 
allowed by the $\rho$ parameter bound lies above both curves. We find that 
for $T=0.18$ and $\mphi =$ 10 GeV, $\vphi$ must be greater than about
440 GeV. On the other hand for $T=0.04$, $\vphi$ must be greater than
about 1000 GeV for the same $\mphi$. The bound on $\vphi$ however
decreases
monotonically with increasing $\mphi$ and becomes about 320 GeV
(for T=0.18)
and 810 GeV (for T=0.04) when $\mphi$ increases to 500 GeV.
The region allowed by the $\rho$ parameter constraint lies above the 
relevant curve.

\newpage

\bfl
{\Large \bf Lower bound on Higgs mass  }
\efl

The beta function for the Higgs self coupling in the presence of a light
radion and the $\rho$ parameter constraints on $m_\phi$ and $\vphi$ can
be used  together to derive a lower bound on $m_h$. 
In the following we briefly 
describe how this lower bound on $m_h$ can be determined.

\vspace*{0.01in}

The beta function for the Higgs self coupling $\lambda$ in the presence of
a light radion is given by {\cite{DM}}

\bea
\beta(\lambda) = \mu \frac{d \lambda}{d \mu} = \frac{1}{8 \pi^2} \left [ 9
\lambda^2 + \frac{402 \lambda^2 v^2}{\vphi^2} + \frac{144 \lambda^2
v^4}{\vphi^4} + \frac{5 \lambda m^2_{\phi}}{\vphi^2} + \lambda \left (6
~g^2_Y - \frac{9}{2}~ g_2^2 - \frac{3}{2}~ g_1^2 \right ) - 6~ g^4_Y 
\right ] \noindent
\nonumber \\
+~\frac{1}{8 \pi^2} \left [\frac{3}{16} \left( g_2^4 + \frac{1}{2}
{(g_2^2 + g_1^2)^2} \right )\right ] 
\eea

\vspace*{0.05in}

i) For a given value of $m_\phi$ we use the above differential equation
to determine the value of the renormalized coupling $\lambda(\mu)$ at $\mu
= 100$ GeV. 
In this paper we shall assume that the Kaluza-Klein modes of the graviton,
which are much heavier than the radion, decouples at
or  above the cut-off scale $\Lambda = 4 \pi \vphi$.
The value of $\lambda$ at the cut-off $\Lambda$ can be
chosen to be either strong and non-perturbative ($\lambda(\Lambda) >
\sqrt{4 \pi}$) or weak and perturbative
($\lambda(\Lambda) < \sqrt{4 \pi}$). In fig. 4 we have plotted 
$\lambda (\mu )$ at $\mu $= 100 GeV against $\vphi$ for seven different
values of $\mphi$ starting from 5 GeV and going up to 600 GeV 
 under the UVBC $\lambda (\L )=\infty $.

ii) For a given $m_\phi$ we find the $\rho$ parameter bound on
$\vphi$ from fig.3. The value of the renormalized coupling
 $\lambda(\mu)$ at that value of $\vphi$
is then determined from the  curve corresponding to the chosen $m_\phi$
shown in fig.4. 
For each chosen $m_\phi$ we therefore obtain a value for
the renormalized coupling $\lambda(\mu)$ at $\mu = $ 100 GeV.
 
\vspace*{0.15in}

iii) The renormalized Higgs mass at $\mu = $ 100 GeV can be determined from the
$\lambda(\mu)$ obtained in step (2) via the relation $m_h(\mu) =
{\sqrt{2 \lambda(\mu)}}~v$. Fig 5 shows the lower bound on the
higgs mass as a function of $\mphi$ for T=0.04 and T=0.18. We find
that for T=0.18 the bound on $m_h$ is greater than the lower bound on
$m_h$ from direct search at LEPII {\cite{J}} provided
$\mphi <$ 600 Gev. This gives rise to two distinct regions in the
$\mphi$ vs $m_h$ plane: RG1 (which is allowed by LEPII but forbidden
by T=0.18) and RG2 (which is allowed by T=0.18 but forbidden by
LEPII). We also find that for T=0.04 the lower bound on $m_h$ is greater
than the LEPII bound for the entire range of values of $\mphi$ relevant for
a light radion. This gives rise to two distinct regions: RG3 (which is 
allowed by LEPII but forbidden by T=0.04) and RG4 (which is allowed 
both by T=0.04 and LEPII). The bound on $m_h$  decreases monotonically
with increasing $\mphi$ due to the following reasons: 
i) the bound on $\vphi$
decreases with increasing $\mphi$ (fig 3) and 
ii) the value of $\lambda (\mu )$
decreases with decreasing $\vphi$ (fig 4).

The results shown in fig 5 were obtained by using the UVBC $\lambda (\L )
=\infty$. It is worthwhile however to investigate the sensitivity of the 
lower bound on $m_h$  to the UVBC on $\lambda$. In fig 6 we have plotted
the bound on $m_h$ for two different UBVC $\lambda (\L )= \infty$
and $\lambda (\L ) =e$. We find that for T=0.18 the bound on $m_h$ is 
insensitive to the UVBC over the entire range of $\mphi$. Howver for
T=0.04 the bound on $m_h$ is somewhat sensitive to the UVBC. To understand 
this feature we would like to refer the reader to {\cite{DM}} where 
it was shown that the value of $\lambda (100 )$ does not depend on the
UVBC provided $\vphi$ is less than 350 GeV. Although the last condition 
is more or less satisfied for T=0.18 it does not hold at all for
T=0.04 over the entire range of vlaues of $\mphi$.

\bfl
{\Large \bf Acknowledgement:} We would like to thank Prof. M. Einhorn,
 Prof. S. Raychaudhuri and Prof. T. Takeuchi for
several useful~ discussions on this paper.
 Uma Mahanta would also like to thank
the Physics Department of IIT Kanpur for its generous
 support and  hospitality while this
work was being done.
\efl

\newpage

%-----------------------------------------------------------------------
\begin{figure}[htb] 
\begin{center} 
\vspace*{5.5in}
      \relax\noindent\hskip -5.4in\relax{\includegraphics{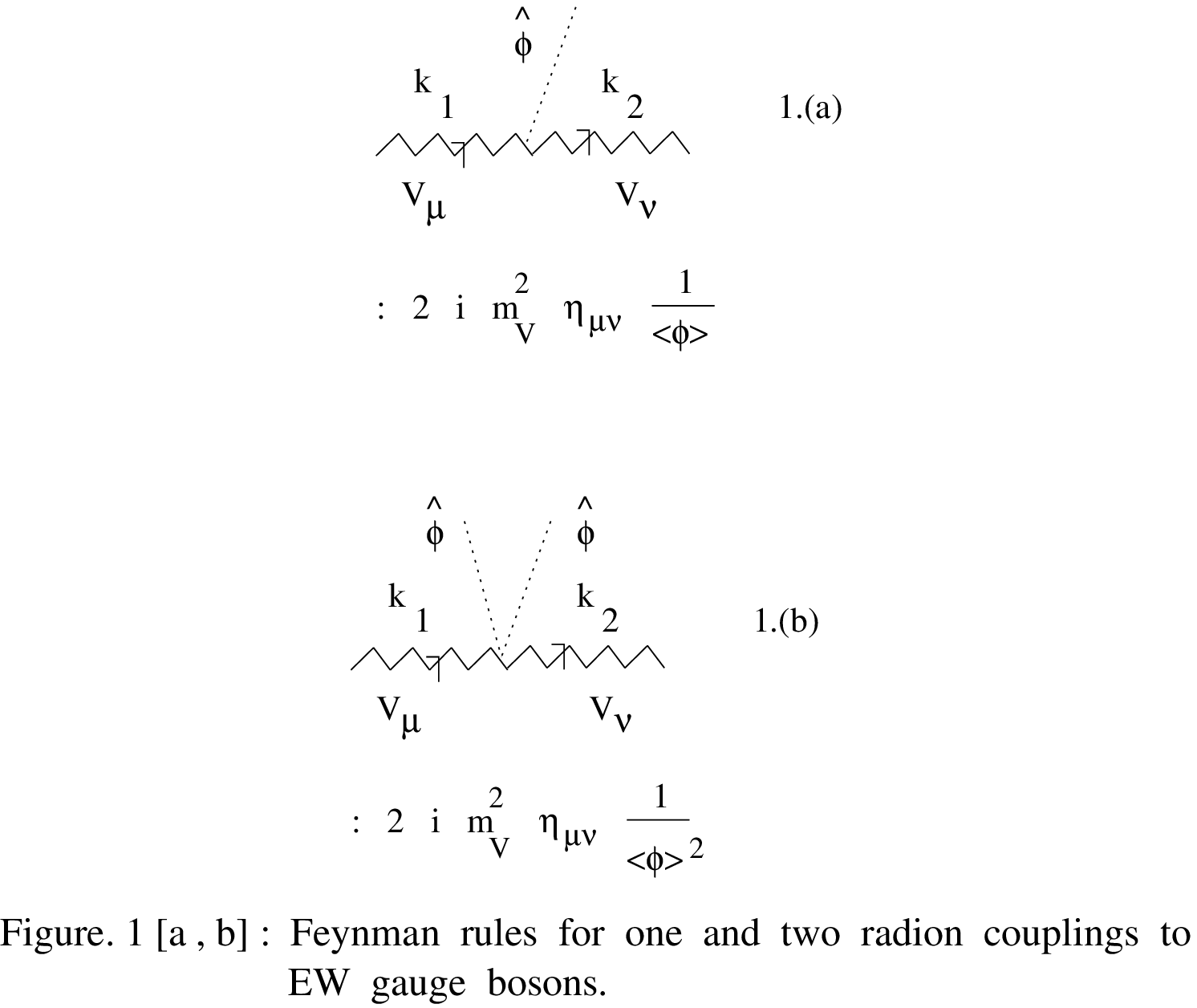}}
\end{center} 
\end{figure} 
%\vspace*{-0.2in} 
%\noindent {\bf Figure. 1[a,b]}.
%{\small { Feynman rules for one and two radion couplings to EW gauge
%bosons.}}
%---------------------------------------------------------------------      

\newpage
%-----------------------------------------------------------------------
\begin{figure}[htb] 
\begin{center} 
\vspace*{2.5in}
      \relax\noindent\hskip -5.7in\relax{\includegraphics{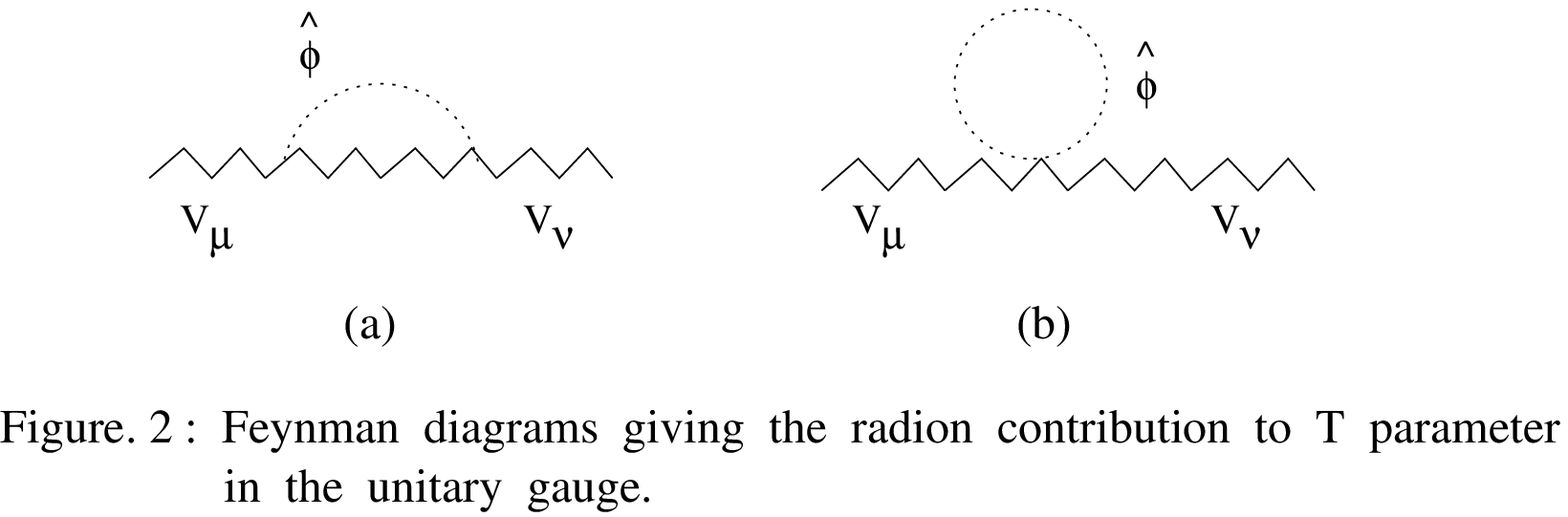}}
\end{center} 
\end{figure} 
%\vspace*{-0.25in} 
%\noindent {\bf Figure. 2}.
%{\small { Feynman diagrams giving the radion contribution to T parameter
%in the unitary gauge.}}
%---------------------------------------------------------------------

\newpage

%-----------------------------------------------------------------------
\begin{figure}[htb] 
\begin{center} 
\vspace*{5.5in}
      \relax\noindent\hskip -5.6in\relax{\includegraphics{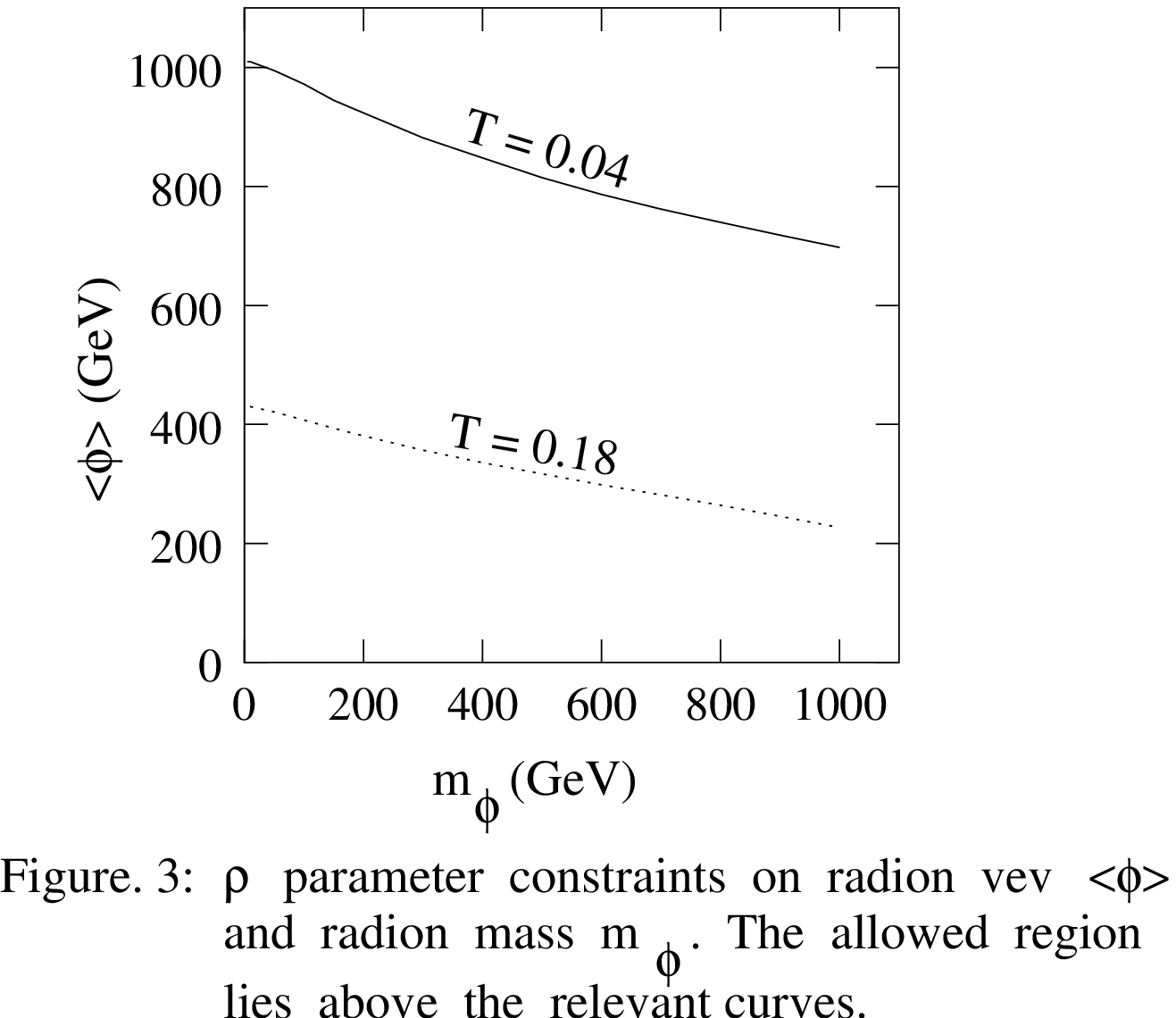}}
\end{center} 
\end{figure} 
%\vspace*{-0.25in} 
%\noindent {\bf Figure. 3}.
%{\small { $\rho$ parameter constraints on radion $\vphi$ and radion mass
%$m_{\phi}$. The allowed region lies above both curves.}}
%---------------------------------------------------------------------
                 
\newpage
 
%-----------------------------------------------------------------------
\begin{figure}[htb] 
\begin{center} 
\vspace*{5.5in}
      \relax\noindent\hskip -5.6in\relax{\includegraphics{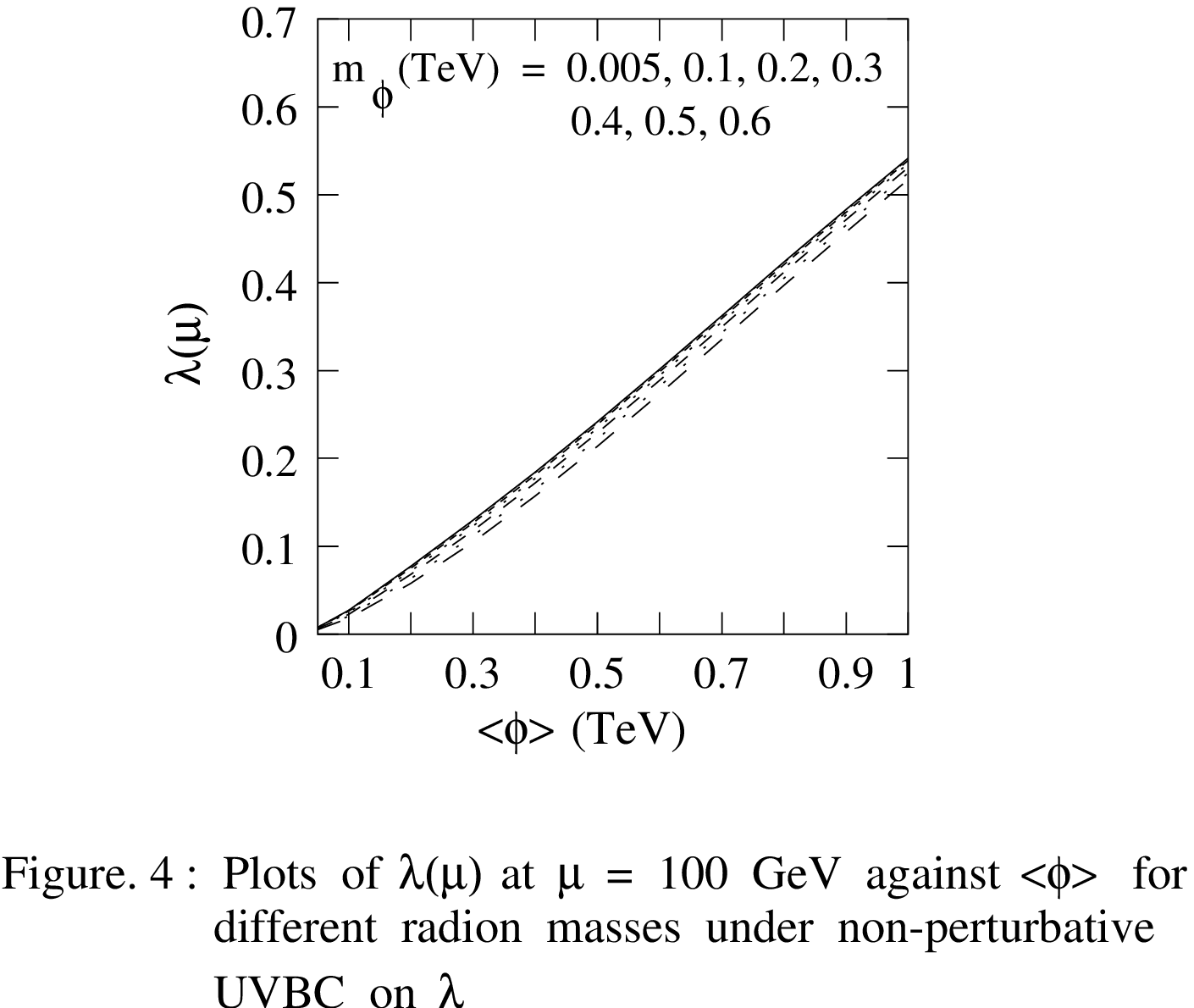}}
\end{center} 
\end{figure} 
%\vspace*{-0.15in} 
%\noindent {\bf Figure. 4}.
%{\small { Plots of $\lambda(\mu)$ at $\mu = 100$ GeV against $\vphi$ for
%different radion masses under non-perturbative UVBC on $\lambda$.}}
% ------------------------------------------------------------------       

\newpage

%-----------------------------------------------------------------------
\begin{figure}[htb] 
\begin{center} 
\vspace*{5.5in}
      \relax\noindent\hskip -6.8in\relax{\includegraphics{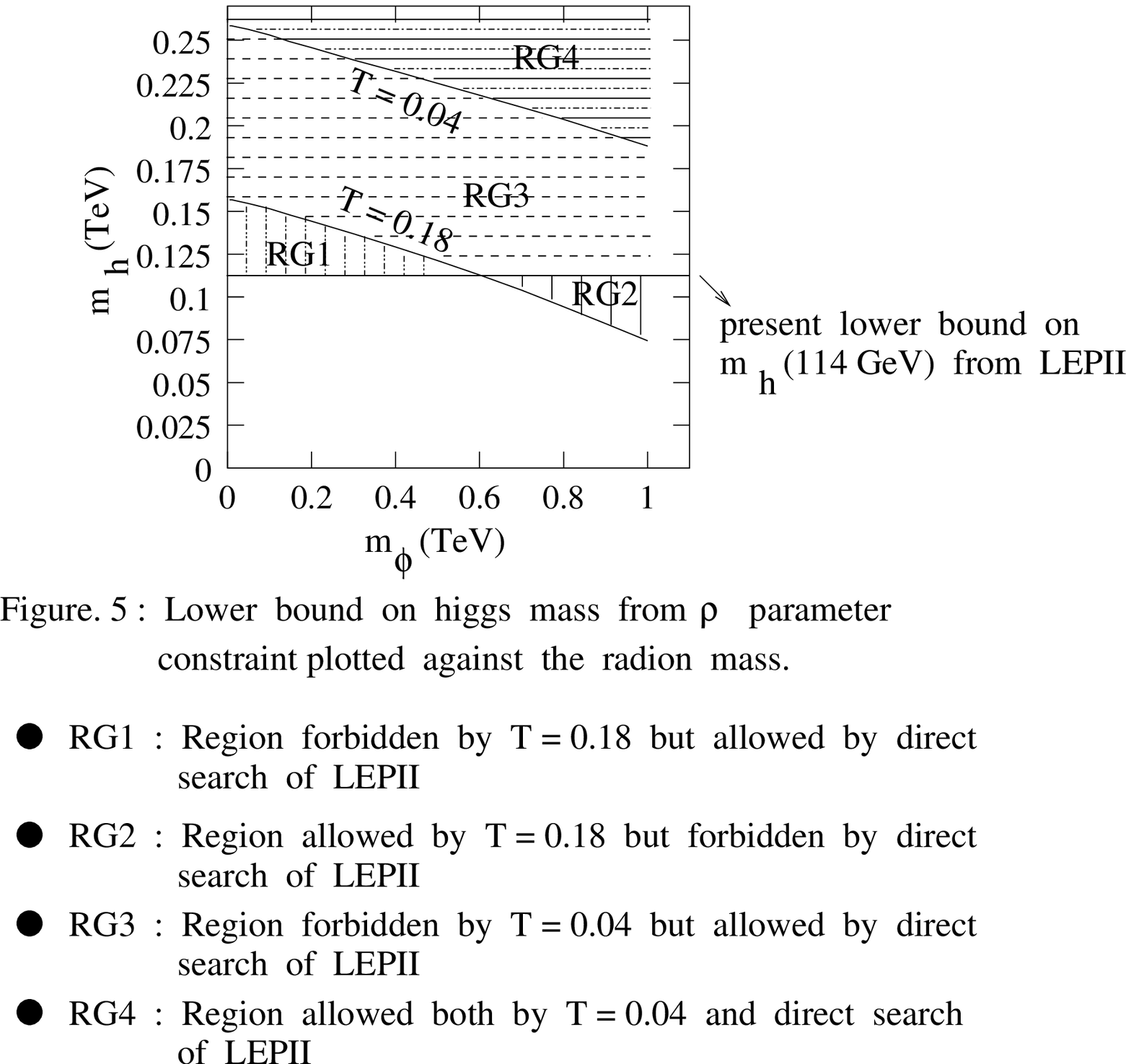}}
\end{center} 
\end{figure} 
%\vspace*{-0.45in} 
%\noindent {\bf Figure. 5}.
%{\small { Lower bound on Higgs mass from $\rho$ parameter constraint
%plotted against the radion mass.}}
% ------------------------------------------------------------------         

\newpage

%-----------------------------------------------------------------------
\begin{figure}[htb] 
\begin{center} 
\vspace*{5.2in}
      \relax\noindent\hskip -6.5in\relax{\includegraphics{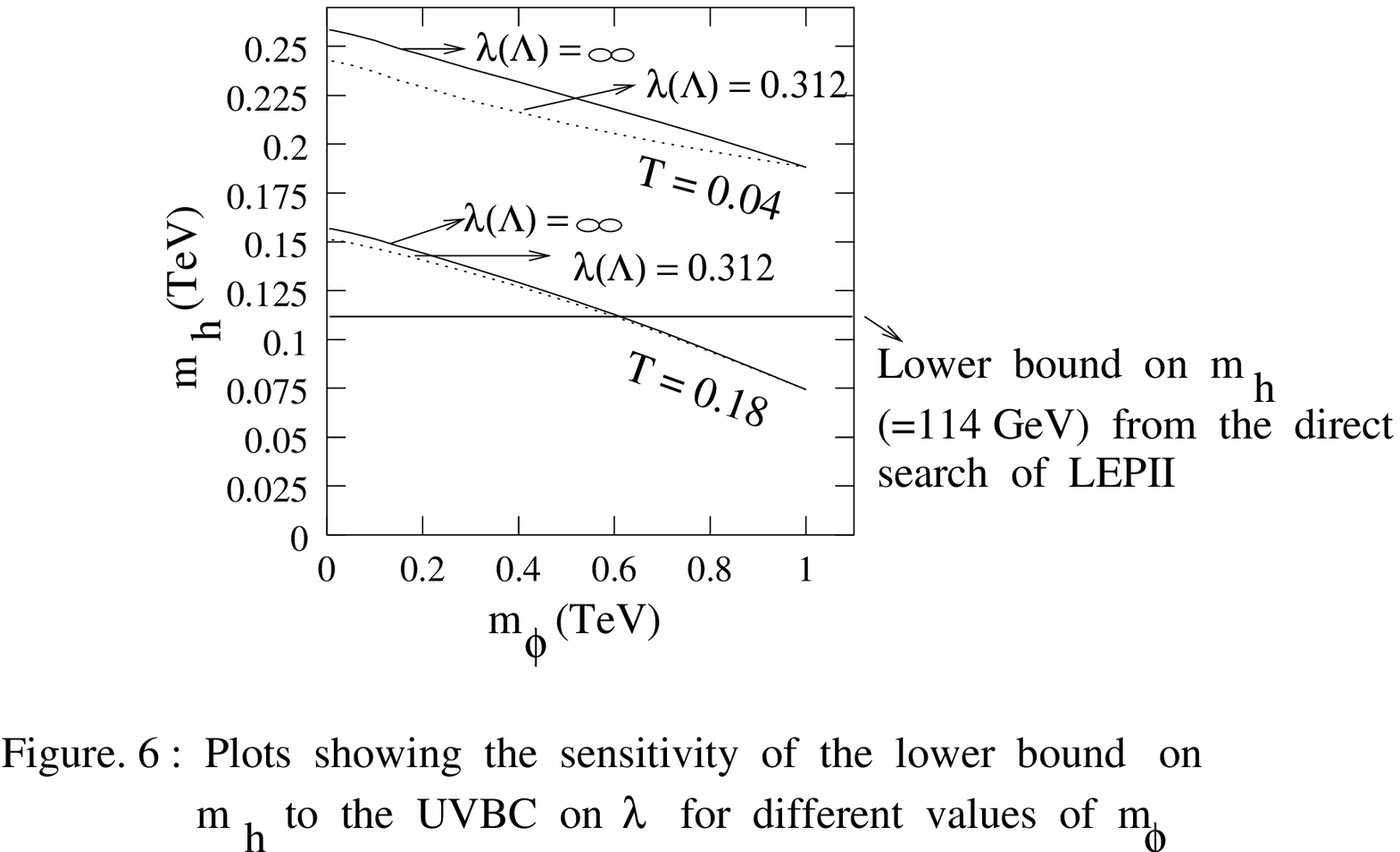}}
\end{center} 
\end{figure} 
%\vspace*{-0.35in} 
%\noindent {\bf Figure. 6}.
%{\small { Plots showing the sensitivity of the lower bound on $m_h$ to
%the UVBC on $\lambda$ for different values of $m_\phi$.}}
%-------------------------------------------------------------------  

\end{document}